\documentclass[reprint, aps, prl]{revtex4-2}
\usepackage{amsmath, amssymb, esint, xspace, braket, color, setspace, accents, silence}

\definecolor{darkred}{rgb}{0.75, 0, 0}
\usepackage[colorlinks=true, linkcolor=darkred, citecolor=darkred, urlcolor=darkred, hypertexnames=false]{hyperref}

\allowdisplaybreaks

\DeclareFontFamily{U}{wasy}{}
\DeclareFontShape{U}{wasy}{m}{n}{
  <-5.5>    wasy5  <5.5-6.5> wasy6  <6.5-7.5> wasy7
  <7.5-8.5> wasy8  <8.5-9.5> wasy9  <9.5->    wasy10
}{}
\DeclareFontShape{U}{wasy}{b}{n}{ <-> wasyb10}{}
\DeclareFontShape{U}{wasy}{bx}{n}{ <-> ssub * wasy/b/n}{}
\DeclareFontShape{U}{wasy}{m}{sl}{ <-> wasysl10}{}
\DeclareFontShape{U}{wasy}{m}{it}{ <-> ssub * wasy/m/sl}{}

\newcommand{\ie}{\textit{i.e.}\@\xspace}
\newcommand{\eg}{\textit{e.g.}\@\xspace}

\newcommand{\etal}{\textit{et al.}\@\xspace}

\newcommand{\eq}[1]{(\ref{#1})}
\newcommand{\msection}[1]{\textit{#1.}---}

\newcommand{\mc}[1]{\mathcal{#1}}
\newcommand{\msf}[1]{\mathsf{#1}}
\newcommand{\mcc}[1]{\mathfrak{#1}}

\newcommand{\m}[1]{\(#1\)}

\newcommand{\favr}[1]{\langle #1 \rangle}

\DeclareMathOperator{\tr}{tr}

\renewcommand{\vec}[1]{\boldsymbol{{#1}}}

\newcommand{\boper}[1]{\hat{\vec{#1}}}

\newcommand{\dd}{\mathrm{d}}
\newcommand{\ee}{\mathrm{e}}
\newcommand{\ii}{\mathrm{i}}
\newcommand{\pd}{\partial}

\newcommand{\eQ}{\vec{\msf{X}}}
\newcommand{\eK}{\vec{\msf{K}}}
\newcommand{\eS}{\vec{\msf{S}}}
\newcommand{\tE}{\tilde{E}}
\newcommand{\tvE}{\tilde{\vec{E}}}

\newcommand{\bdop}[1]{\accentset{\leftrightarrow}{#1}}
\newcommand{\bdopl}[1]{\accentset{\leftarrow}{#1}}
\newcommand{\bdopr}[1]{\accentset{\rightarrow}{#1}}
\newcommand{\ldX}{\accentset{\leftarrow}{\pd}_{x}}
\newcommand{\ldP}{\accentset{\leftarrow}{\pd}_{k_x}}
\newcommand{\rdX}{\accentset{\rightarrow}{\pd}_{x}}
\newcommand{\rdP}{\accentset{\rightarrow}{\pd}_{k_x}}

\begin{document}

\title{Resonant and Ponderomotive Pumping of Zonal Flows by Turbulence, {Alfv\'en} Eigenmodes, and Radiofrequency Waves}

\author{C. M. Gallaro}
\email{cg4025@princeton.edu}
\affiliation{Department of Astrophysical Sciences, Princeton University, Princeton, NJ 08544, USA}

\author{I. Y. Dodin}
\email{idodin@sjtu.edu.cn}
\affiliation{Future Energy Institute and School of Engineering, Shanghai Jiao Tong University, Shanghai 200240, China}
\affiliation{Princeton Plasma Physics Laboratory, Princeton, NJ 08543, USA}

\date{\today}

\begin{abstract}
Analytical studies of zonal-flow (ZF) excitation by fluctuating fields, such as drift-wave turbulence and Alfv\'en eigenmodes (AEs) driven by energetic particles (EPs) in tokamaks, are usually carried out via intricate gyrokinetic calculations and simulations whose results are notoriously difficult to interpret. Here, we report a transparent result that is also not limited to any particular modes or frequency range. Using oscillation-center theory that captures both ponderomotive forces and quasilinear diffusion, we derive a compact formula for the ZF drive produced by any fluctuating field in terms of the field's canonical momentum and the dissipation power density. For ZFs generated by drift waves, our model subsumes the local relation between the zonal velocity and the drift-wave energy density that was previously derived ad hoc. The simplicity and generality of our result also opens a path toward optimization of ZF excitation with external waves and bridges the physics of AE--EP interactions with that of poloidal flows driven by radiofrequency waves in fusion plasmas.
\end{abstract}

\maketitle

\msection{Introduction} Zonal flows (ZFs) play a critical role in the nonlinear self-organization of turbulence in magnetized fusion plasmas \cite{ref:rogers00, ref:diamond05, ref:zhu19}. Lately, a lot of attention has been drawn to this topic in connection with the confinement of energetic particles (EPs) in such plasmas \cite{ref:qiu16, ref:disiena21, ref:choi23}, which is mainly determined by resonant interactions with Alfv\'en eigenmodes (AEs) \cite{ref:gorelenkov18, ref:gorelenkov19, ref:gorelenkov24} and scattering caused by drift-wave (DW) turbulence \cite{ref:chen16, ref:duarte17}. It is understood now that ZFs can play a significant role in this process in that they both reduce AEs' energy \cite{ref:chen25, ref:biancalani20, ref:yan25} and suppress turbulence \cite{ref:lin98, ref:sama24, ref:du25, ref:garcia24, ref:ruiz25}, thereby reducing EP transport \cite{ref:brochard24, ref:brochard242, ref:barberis25}. It is therefore important to understand how ZF excitation works in this system.

Commonly known are two main channels via which such excitation can occur: (i) ZFs can be driven by the ponderomotive forces produced by AEs and (ii) they can also be generated directly by EPs, whose spatial quasilinear (QL) diffusion causes charge separation between flux surfaces and thereby drives \m{E \times B} flows. Much effort has been invested in studies of these effects, mostly based on gyrokinetic theory and simulations \cite{ref:chen12, ref:chen16, ref:qiu16, ref:qiu17, ref:chen22, ref:chen25}. However, the results of these studies remain notoriously difficult to interpret, so there has been a demand for a more transparent description of the ZF drive caused by both adiabatic and resonant interactions.

Here, we propose such a description based on oscillation-center (OC) quasilinear theory (QLT). OC QLT was originally developed in \cite{ref:dewar73, ref:mcdonald85} and represents a revision of classic QLT \cite{book:stix}. It turns the latter into a local theory applicable to inhomogeneous plasmas and captures both resonant diffusion and ponderomotive forces, which makes it a perfect tool for the job. As a bonus, OC QLT is not limited to low-frequency waves and thus potentially allows for more general results than gyrokinetics. We apply the recent formulation of OC QLT \cite{my:ql, my:qlrmpp} to derive a compact equation for the zonal acceleration of a plasma driven by a general fluctuating field. This field can be a monochromatic wave or broadband turbulence and is not limited to any particular polarization or frequency range. It also does not have to satisfy a dispersion relation and can be externally imposed. We show that our result subsumes the local relation between the zonal velocity and DW's energy density that was found ad hoc in \cite{ref:zhou19, ref:zhu21}. The simplicity and generality of our result also opens a path toward optimization of ZF excitation with external waves and bridges the physics of AE--EP interactions with that of poloidal flows driven by radiofrequency (RF) waves in fusion plasmas \cite{ref:berry99, ref:batchelor99, ref:LeBlanc99, ref:elf00, ref:lin08, ref:guan}. The model we present here is limited to slab geometry, but general geometry can also be accommodated within OC QLT, as to be discussed in future work.

\msection{A primer on OC QLT} Consider some plasma species interacting with a fluctuating field \m{\tilde{\vec{E}}(t, \vec{x})} with a characteristic amplitude \m{\tE}. Assume that the particle Hamiltonian has the form \m{H \approx H_0 + H_1 + H_2}, where \m{H_n} scales as \m{\tE^n}. Then, under the assumptions specified below, the slow dynamics of these species can be described by the OC kinetic equation \cite{my:ql, my:qlrmpp}
\begin{gather}\label{QLevolution}
\frac{\pd F}{\pd t} - \left\{\mc{H}, F\right\}  = \frac{\pd}{\pd \vec{P}}\cdot \left(\vec{D} \, \frac{\pd F}{\pd \vec{P}}\right)+\msf{C}.
\end{gather}
The function \m{F} is the OC (``dressed'') distribution:
\begin{gather}\label{DressedDist}
F \doteq \favr{f} + \frac{1}{2} \frac{\pd}{\pd \vec{P}}\cdot \left(\vec{\Theta}\,  \frac{\pd \favr{f}}{\pd \vec{P}}\right),
\end{gather}
where \m{\doteq} denotes a definition. The statistical properties of the system are assumed weakly, if at all, inhomogeneous in time \m{t} and in some generalized coordinates \m{\vec{Q}}. The Poisson bracket in \eq{QLevolution} can be calculated in any phase-space variables, but the momenta \m{\vec{P}} must be canonically conjugate to \m{\vec{Q}}. The function \m{f} is the particle phase-space probability distribution, and the angular brackets denote local averaging over the field fluctuations. The matrix \m{\vec{\Theta}} is the dressing function, \m{\mc{H} = H_0 + \Delta} is the OC Hamiltonian, \m{\Delta} is the ponderomotive energy, and \m{\vec{D}} is the diffusion matrix. These functions are defined as follows. Consider \m{H_1} as a field on the extended phase space \m{\eQ \doteq (t, \vec{Q}, \vec{P})} of dimension \m{\msf{n} \doteq \dim \eQ}. Associated with it is a two-point correlation function \m{C(\eQ, \eS) \doteq \favr{H_1(\eQ + \eS/2)\,H_1(\eQ - \eS/2)}} and the Fourier image thereof, \m{W(\eQ, \eK) \doteq (2\pi)^{-\msf{n}}\int C(\eQ, \eK)\,\ee^{-\ii \eK \cdot \eS}\,\dd\eS}, also known as the (average) Wigner function of \m{H_1}. Assuming the notation \m{\eK \equiv (\omega, \vec{K}, \vec{R})}, one has
\begin{subequations}\label{eq:nlp}
\begin{align}
\vec{\Theta} 
& \doteq \frac{\pd}{\pd \vartheta} \fint \dd\eK\,\frac{\vec{K}\vec{K} W(\eQ, \eK)}{\omega - \vec{K}\cdot\vec{V} + \vartheta}\bigg|_{\vartheta = 0},
\\
\Delta 
& \doteq \favr{H_2} + \frac{1}{2}\frac{\pd}{\pd \vec{P}} \cdot 
\fint \dd\eK\,\frac{\vec{K} W(\eQ, \eK)}{\omega - \vec{K}\cdot\vec{V}},
\\
\vec{D} 
& \doteq \pi \int \dd\eK\,\vec{K}\vec{K}\,
\delta(\omega - \vec{K}\cdot \vec{V}) W(\eQ, \eK),
\end{align}
\end{subequations}
where \m{\vec{V} \doteq \pd_{\vec{P}}\mc{H}} is the OC velocity and \m{\vec{K}\vec{K}} is a dyadic. The first two integrals, denoted \m{\fint}, are Cauchy principal values and thus do not diverge at wave--particle resonances. (Other than that, integrals in this paper are, by default, taken over the real line \m{(-\infty, \infty)} in each dimension.) The term \m{\msf{C}} in \eq{QLevolution} is a generalized Balescu--Lenard collision operator and is neglected below. 

The first-order interaction Hamiltonian can be expressed as \m{H_1 = \smash{\boper{\alpha}}^\dag\tvE}, where \m{\boper{\alpha}} is a linear operator acting in the space of functions of \m{\eQ}. (In \cite{my:ql}, it is assumed local in \m{\vec{P}}, but waiving this assumption does not change the result.) Then, one can express \m{W} through the (Weyl) symbol of \m{\smash{\boper{\alpha}}} and the Wigner matrix of \m{\tvE(t, \vec{x})},
\begin{multline}
\mcc{W}_{ab}(t, \vec{x}, \omega, \vec{k}) \doteq \int \frac{\dd\tau}{2\pi}\,\frac{\dd\vec{s}}{(2\pi)^{n_x}}\,
\ee^{\ii\omega \tau - \ii \vec{k} \cdot \vec{s}}\\
\times 
\favr{
\tilde{E}_a(t + \tau/2, \vec{x} + \vec{s}/2)
\tilde{E}_b(t - \tau/2, \vec{x} - \vec{s}/2)
},\label{eq:Mcc}
\end{multline}
where \m{n_x \doteq \dim\vec{x}}. For geometrical-optics (GO) waves, one can couple OC QLT with the wave-kinetic equation for the wave-action density (in the ray phase space), a functional of \m{\vec{\mcc{W}}}. This yields a local theory that conserves the energy--momentum of the wave--plasma system \textit{and} the actions of nonresonant waves \cite{ref:dodin22, ref:dodin24}.

\msection{OC QLT for magnetized plasma} Application of this framework to magnetized plasmas will be detailed in a separate paper (Paper~II) and is only outlined here. We assume \m{H(t, \vec{x}, \vec{p}) = (\vec{p} - q\vec{A}(\vec{x})/c - q\tilde{\vec{A}}(t, \vec{x})/c)^2/2m + q \varphi(t, \vec{x})}, where \m{\vec{x}} is the Euclidean spatial coordinate (\m{n_x = 3}), \m{\vec{p}} is the canonical momentum, \m{q} and \m{m} are the particle charge and mass, and \m{c} is the speed of light. The vector potential \m{\vec{A}} determines the background magnetic field \m{\vec{B} = \nabla \times \vec{A} \equiv \vec{b}B}, which is assumed stationary. The vector potential \m{\tilde{\vec{A}}} describes electromagnetic (or electrostatic) fields that fluctuate rapidly compared to the ZF evolution. For example, \m{\tilde{\vec{A}}} can describe AEs, DWs, or other waves. The Weyl gauge for the fluctuating field is adopted, so \m{\tvE = - \pd_t \tilde{\vec{A}}/c}. The electrostatic potential \m{\varphi} describes only slow fields, such as ZFs.

Having order-one \m{\vec{B}} implies that spatial derivatives of \m{\vec{A}} are not small, so adopting \m{(\vec{Q}, \vec{P}) = (\vec{x}, \vec{p})} is not an option. Instead, one can use angle--action coordinates or a local approximation thereof, for example, those constructed in \cite{ref:wong00}. But since our results are not significantly affected by inhomogeneity of \m{\vec{B}} (except, possibly, for trapped-particle effects, which we neglect), we assume constant \m{\vec{B}}, with \m{\vec{A} = B x \vec{e}_y}, where \m{\vec{e}_i} is the unit vector along the \m{i}th axis. (In a tokamak, \m{x} and \m{y} would be the radial and poloidal coordinates, respectively, and \m{z} would be the coordinate along \m{\vec{B}}; \ie \m{\vec{e}_z \equiv \vec{b}}.) Then, we introduce canonical gyrocenter coordinates \m{\vec{Q} = (\phi, Y, Z)}, where \m{\phi} is the gyrophase, and the conjugate momenta \m{\vec{P} = (P_\phi, P_Y, P_Z)}. (The construction of these coordinates is explained in Paper II.) With \m{X \doteq P_Y/m\Omega}, where \m{\Omega \doteq qB/mc}, \m{\vec{X} \doteq (X, Y, Z)} serves as the gyrocenter spatial location. This leads to \m{H_0 = P_Z^2/2m - \Omega P_\phi + q \favr{\varphi}_{\text{gy}}}, \m{H_1(t, \vec{x}, \vec{p}) = \smash{\boper{\alpha}}^\dag\tvE(t, \vec{X})}, and \m{\favr{H_2} = q^2\favr{\tilde{A}^2}/2mc^2}. Here, \m{\favr{\varphi}_{\text{gy}}} is the gyroaverage of \m{\varphi}, \m{\boper{\alpha} = \sum_{l = -\infty}^\infty \ee^{\ii l\phi}\boper{\alpha}_l}, and \m{\boper{\alpha}_l} can be inferred from \cite{ref:grebogi79}. Then, for any relevant \m{\eK}-dependent function \m{\Psi}, one has \cite{tex:supp}
\begin{gather}
\int \dd\eK\,\Psi W = \sum_l \int \dd\omega\,\dd\vec{k}\,\Psi_l(\omega, \vec{k}) \msf{W}_l(t, \vec{x}, \omega, \vec{k}),
\end{gather}
where \m{\Psi_l \doteq \Psi(K_\phi = l)}, \m{\msf{W}_l = \tr(\vec{\alpha}_l^\dag \star \vec{\mcc{W}} \star \vec{\alpha}_l)}, \m{\tr} is trace, \m{\vec{\alpha}_l} are the symbols of \m{\boper{\alpha}_l}, and \m{\star} is the Moyal product. Using \m{\bdopl{\vec{\alpha}}_l \doteq \star \vec{\alpha}_l} and \m{\bdopr{\vec{\alpha}}_l^\dag \doteq \vec{\alpha}_l^\dag \star}, one can rewrite this as
\begin{gather}
\msf{W}_l = \tr(\bdopr{\vec{\alpha}}_l^\dag \vec{\mcc{W}} \bdopl{\vec{\alpha}}_l)
\equiv \vec{\mcc{W}} : \bdopl{\vec{\alpha}}_l\bdopr{\vec{\alpha}}_l^\dag,
\label{eq:dar}
\end{gather}
where the second equality \textit{defines} the operation denoted with colon (a generalized trace). As usual, the arrows indicate the directions in which the derivatives act.

Per OC QLT assumptions, we require the GO approximation in \m{t} and in the canonical coordinates \m{Y} and \m{Z}; \ie \m{F(t, \vec{x}, \vec{p})} and \m{\vec{\mcc{W}}(t, \vec{x}, \omega, \vec{k})} must be slow in \m{(t, y, z)} compared to the corresponding oscillation scales. (Uppercase and lowercase arguments are used interchangeably where they serve as placeholders.) No such assumption is made about the \m{x}-direction: since \m{m \Omega X} is a canonical \textit{momentum}, order-one \m{\pd_x F} and \m{\pd_x\vec{\mcc{W}}} are allowed. Then, \m{\bdopl{\vec{\alpha}}_l \approx \exp(\frac{\ii}{2}\,\ldX \rdP)\vec{\alpha}_l} and \m{\bdopl{\vec{\alpha}}_l^\dag \approx \vec{\alpha}_l^\dag\exp(-\frac{\ii}{2}\,\ldP \rdX)}. By a known theorem \cite{phd:ruiz17}, one has \m{\int \dd x\,\dd k_x \msf{W}_l = \tr \int \dd x\,\dd k_x (\vec{\alpha}_l^\dag \vec{\mcc{W}} \vec{\alpha}_l)}. If \m{\vec{\mcc{W}} \propto \delta(x - x_c)}, as in a radially localized eigenmode, then
\begin{gather}
\int \dd x\,\dd k_x \psi \msf{W}_l \approx 
\psi(x_c) \tr \int \dd x\,\dd k_x\, (\vec{\alpha}_l^\dag \vec{\mcc{W}} \vec{\alpha}_l),
\end{gather}
for any \m{\psi} much wider than \m{\vec{\mcc{W}}}. In the opposite limit, if \m{\vec{\mcc{W}}} is statistically homogeneous in a wide (many wavelengths across) region in \m{x}, one has \m{
\msf{W}_l \approx \tr (\vec{\alpha}_l^\dag \favr{\vec{\mcc{W}}}_x \vec{\alpha}_l)}, where \m{\favr{\vec{\mcc{W}}}_x} is the local \m{x}-average of \m{\vec{\mcc{W}}}.

\msection{OC density equation} By integrating \eq{QLevolution} over \m{\dd\vec{p}} (now, \m{\vec{p}} is the kinetic momentum), one obtains an equation for the OC density \m{N (t, \vec{x}) \doteq \int \dd\vec{p}\,F (t, \vec{x}, \vec{p})} \cite{tex:supp}:
\begin{gather}\label{eq:N}
\pd_t N + \nabla \cdot (N \vec{U}) + (\pd_x \mc{Q})/(m \Omega) = 0,
\end{gather}
where \m{\vec{U}} is the OC fluid velocity and \m{\mc{Q}  \doteq -   \int \dd \vec{p}\, D_{yj} \pd_{P_j} F}. Assuming the flows are nonlinearly driven by \m{\tvE}, one has \m{\vec{U}   = \mc{O}(\tilde{E}^2)} or \m{\vec{U}   = \mc{O}(\tilde{E})}, if \m{\vec{U}} is amplified by a modulational instability during \m{t = (\tilde{E}^{-1})}; in either case, \m{N  \vec{U}  \approx n  \vec{U}}. Here, \m{n (t, \vec{x}) \doteq \int \dd\vec{p}\, \favr{f}(t, \vec{x}, \vec{p})} is the average (over \m{\tvE} oscillations) gyrocenter density, and we neglected terms smaller than \m{\mc{O}(\tilde{E}^2)}. From \eq{DressedDist}, one can show that \m{N  = n  + (\pd_x\mc{P})/m \Omega}, where \m{\mc{P}  \doteq \int \dd \vec{p}\, \Theta_{Yj} \pd_{P_j} \favr{f}/2}. Also \cite{tex:supp},
\begin{subequations}\label{eq:QPmcc}
\begin{gather}
\mc{P} = \int \dd\omega\,\dd\vec{k}\,\frac{k_y}{\omega}\,\frac{\vec{\mcc{W}}(\omega, \vec{k})}{8\pi\omega}: \,\frac{\pd (\omega^2\bdop{\vec{\chi}}_{\text{H}}(\omega, \vec{k}))}{\pd\omega},
\\
\mc{Q} = \int \dd\omega\,\dd\vec{k}\,\frac{k_y}{4\pi}\, \vec{\mcc{W}}(\omega, \vec{k}) : \bdop{\vec{\chi}}_{\text{A}}(\omega, \vec{k}),
\label{eq:Qs}
\\
 \bdop{\vec{\chi}} \doteq \sum_{l} \int \dd \vec{p} \, \frac{4\pi \bdopl{\vec{\alpha}}_{l}(\vec{K}_l \cdot \pd_{\vec{P}} F)\bdopr{\vec{\alpha}}_{l}^\dag}{\omega + l\Omega  - k_z p_z/m + \ii 0} - \frac{\omega_p^2}{\omega^2}\,\vec{1},
 \label{eq:chis}
\end{gather}
\end{subequations}
where \m{\vec{K}_l \doteq (l, k_y, k_z)}, \m{\vec{P} = {(p_\phi, m\Omega x, p_z)}}, and \m{\omega_p} is the plasma frequency. Slow dependence on \m{(t, \vec{x})} is not emphasized but allowed for all functions. The indices~\m{\text{H}} and~\m{\text{A}} denote the principal-value part, plus \m{-(\omega_p^2/\omega^2)\vec{1}}, and the resonant-pole part of the integral \eq{eq:chis}.

For example, for a quasimonochromatic wave with frequency \m{\omega}, wavevector \m{\vec{k}}, and envelope \m{\tilde{\vec{E}}_0}, one has~\cite{tex:supp}
\begin{subequations}\label{eq:QP}
\begin{gather}
\favr{\mc{P}}_x = \frac{k_y}{16\pi \omega^2}\,  \smash{\tilde{\vec{E}}}^\dag_0 \, \frac{\pd}{\pd \omega}\left(\omega^2 \vec{\chi}_{\text{H}}(\omega, \vec{k}) \right) \tilde{\vec{E}}_0,
\label{eq:Pim}
\\
\favr{\mc{Q}}_x =  \frac{k_y}{8\pi}\,\tilde{\vec{E}}^\dag_0 \vec{\chi}_{\text{A}}(\omega, \vec{k})\tilde{\vec{E}}_0.
\label{eq:Qss}
\end{gather}
\end{subequations}
Here, \m{\vec{\chi}_{\text{H}}} and \m{\vec{\chi}_{\text{A}}} are the Hermitian and anti-Hermitian parts of \m{\vec{\chi}} that is given by \eq{eq:chis} without the arrows. One can also show (Paper~II) that said \m{\vec{\chi}} is precisely the GO susceptibility matrix, \m{\vec{\chi} = \vec{\chi}_0 + \delta \vec{\chi}}, where \m{\vec{\chi}_0} is the susceptibility of given species in a homogeneous magnetized plasma, as in \cite[Chap.~10]{book:stix}, and \m{\delta \vec{\chi}} is the correction caused by non-negligible \m{\pd_x F}, as in \cite[Sec.~14.6]{book:stix}. Then, one can recognize \m{\favr{\mc{P}}_x} as field's canonical poloidal momentum stored in the OC dressing; cf.\ \cite[(132) and (133)]{my:amc}. (The \textit{total} momentum of the field is the sum of the material momentum carried by all species plus the vacuum contribution \m{k_y |\tilde{E}_0|^2/8\pi\omega}. The latter is negligibly small when the total susceptibility is large, \eg for AEs and DWs.) Likewise, one can recognize \m{\favr{\mc{Q}}_x} as the loss of field's poloidal canonical momentum, to a given species, per unit volume per unit time, \m{\favr{\mc{Q}}_x  = k_y \msf{P}_{\text{abs}}/\omega}, where \m{\msf{P}_{\text{abs}}} is the field energy dissipated collisionlessly, to a given species, per unit volume per unit time \cite{book:stix}. Equations \eq{eq:QPmcc} only generalize these formulas to broad spectra and possibly lack of scale separation in the \m{x}-direction. 

\msection{Zonal drive} Let us now consider multiple species \m{s} and calculate the average flow produced by an unspecified oscillating field in the zonal direction, which in our geometry is the \m{y}-axis. We start by rewriting \eq{eq:N} as an equation for the gyrocenter density \m{n_s}:
\begin{gather}
\pd_t (m_s\Omega_s n_s + \pd_x \mc{P}_s)  + \nabla \cdot (m_s\Omega_s n_s \vec{U}_s) + \pd_x \mc{Q}_s = 0.
\notag
\end{gather}
The flow velocity \m{\vec{U}_s} is a sum of the \m{E \times B} drift \m{\vec{U}_E = (c/B)\vec{b} \times \nabla \varphi} and the ponderomotive drift \m{\vec{U}_{\Delta_s} = (c/q_sB)\vec{b} \times \nabla \Delta_s}. (The latter is negligible at low frequencies \cite{tex:supp}.) Let us sum the above equation over species and use the gyrocenter Poisson's equation for the low-frequency dynamics, \m{\nabla_\perp \cdot (\chi_\perp \nabla_\perp \varphi) = - 4\pi \sum_s q_s n_s}. Here, \m{_\perp} denotes the plane transverse to \m{\vec{B}}, \m{\chi_\perp = c^2/V_{\text{A}}^2} is plasma's low-frequency transverse susceptibility, \m{V_{\text{A}} = B/\sqrt{4\pi\varrho} \ll c} is the Alfv\'en speed, and \m{\varrho \doteq \sum_s m_s n_s} is the plasma mass density. It is easy to see that \m{\sum_s m_s\Omega_s n_s = - \nabla_\perp \cdot (\vec{\Pi} \times \vec{b})}, where \m{\vec{\Pi} \doteq \varrho c\vec{b} \times \nabla\varphi/B} is the \m{E \times B} momentum density. Then, after applying poloidal averaging, denoted with a bar, one obtains \cite{tex:supp}
\begin{gather}
\pd_t \pd_x \overline{\Pi}_y = 
\pd_t\pd_x \overline{\mc{P}} + \pd_x \overline{\mc{Q}} +
\sum_s m_s\Omega_s\overline{\nabla \cdot (n_s \vec{U}_s)},
\end{gather}
where \m{\overline{\mc{P}} \doteq \sum_s \overline{\mc{P}}_s} represents the total material momentum density and \m{\overline{\mc{Q}} \doteq \sum_s \overline{\mc{Q}}_s} represents the total loss of field's poloidal canonical momentum per unit volume per unit time. From the fact that \m{n_s} and \m{\vec{U}_s} are defined as slow quantities and \m{\vec{U}_s \propto \vec{b} \times \nabla (\ldots)}, it is easy to see \cite{tex:supp} that the last term is negligible. Then, assuming a simplified notation \m{\Pi \equiv \overline{\Pi}_y} and omitting the other bars too from now on, one obtains
\begin{gather}\label{eq:zd}
\pd_t \Pi = \pd_t \mc{P} + \mc{Q},
\end{gather}
where \m{\bdop{\vec{\chi}}} are now calculated on poloidally averaged \m{n_s}.

Equation \eq{eq:zd} is the main result of this paper. It is most easily understood when, in \eq{eq:QPmcc}, the arrows can be omitted and the colons can be replaced with a regular trace. This applies to: (i) the local \m{x}-average of \eq{eq:zd}, when \m{\favr{\vec{\mcc{W}}}_x} is well defined (\ie the GO approximations holds for the \m{x}-axis as well); (ii) the \m{x}-integrated \eq{eq:zd}, \m{\pd_t \int \Pi\,\dd x = \pd_t \int \mc{P}\,\dd x + \int \mc{Q}\,\dd x}, when waves are localized in \m{x} on a scale much smaller than the scale of all \m{F_s} that matter. In any case, \m{\mc{Q}} represents resonant transfer of the canonical momentum from waves to particles. (For AEs, it is commonly identified as CCT \citep{ref:qiu17}.) The presence of \m{\pd_t\mc{P}} reflects the fact that the material momentum is simply the average kinetic momentum of the background plasma \cite{my:ql, my:sharm, my:amc}, which, in our case, is the zonal momentum. This term describes the cumulative effect of the Reynolds and Maxwell stresses produced by \m{\tilde{\vec{A}}} \cite{ref:gao06, ref:gao07}, as well as the so-called diamagnetic stress \citep{ref:zhang26}. 

\msection{Example: DWs} As an example, consider DWs within the modified Hasegawa--Mima model \cite{ref:zhu21}, which assumes cold ions (\m{s = i}) with mass \m{m_i} and hot massless electrons (\m{s = e}) with constant temperature \m{T_e}. It is assumed that \m{q_i = - q_e \equiv e}, so \m{n_e \approx n_i \equiv n}, and the sound speed is \m{c_{\text{S}} \doteq \sqrt{T_e/m_i}}. It is also assumed that \m{\rho_{\text{S}} \doteq c_{\text{S}}/\Omega_i} is comparable to the transverse wavelength \m{k_\perp^{-1}}, while both \m{\rho_e} and \m{\rho_i} are negligible. Then, one obtains \m{\vec{\chi} = (\vec{1} - \vec{b}\vec{b}^\dag)\chi_\perp + \vec{b}\vec{b}^\dag(k_z \lambda_{\text{D}e})^{-2}(1 - \omega_*/\omega)}, where \m{\omega_* = k_y U_*} emerges via \m{\delta\vec{\chi}}; cf.\ \cite[Secs.~3.4-3.8]{book:stix}. Here, \m{\lambda_{\text{D}e} = \sqrt{T_e/4\pi n e^2}} is the electron Debye length, \m{U_* = c T_e/e B L} is the diamagnetic drift velocity, and \m{L^{-1} \doteq - \pd_x \ln n}. DWs satisfy the dispersion relation \m{\omega = \omega_*/(1 + k_\perp^2 \rho_{\text{S}}^2)} and collisionless dissipation is absent, so \m{\mc{Q} = 0}. Then, \eq{eq:zd} yields \cite{tex:supp}
\begin{gather}\label{eq:Mdw}
\favr{\Pi}_x =\favr{\favr{\tilde{w}^2}}_x/8\pi U_*\lambda_{\text{D}e}^2,
\quad
\tilde{w} \doteq (\rho_{\text{S}}^2\nabla_\perp^2 - 1)\tilde{\varphi}.
\end{gather}
(The double brackets mean averaging over \m{(t, y, z)} \textit{and} over \m{x}.) This agrees \cite{tex:supp} with the result that was obtained in \cite{ref:zhou19, ref:zhu21} (see also \citep{ref:chen24}) from other considerations.

\msection{Example: AEs} Assuming the cold-plasma limit and \m{\chi_\perp \gg 1}, one can neglect the parallel field and use just the transverse susceptibility, \m{\vec{\chi} \approx (\vec{1} - \vec{b}\vec{b}^\dag)\chi_\perp}. Then, to the extent that collisionless dissipation is negligible, \eq{eq:zd} leads to \m{\Pi \sim k_y |\tilde{B}|^2/8\pi\omega}, assuming a mode with given \m{\omega} and \m{k_y}. For example, consider DIII-D parameters, say, \m{B \sim 2.1\,\text{T}}, \m{n \sim 10^{13}\,\text{cm}^{-3}}, \m{\tilde{B}/B \sim 10^{-3}}, \m{q \sim 5}, at \m{R\sim 2.1\,\text{m}}, \m{k_y \sim 0.5\,\text{cm}^{-1}}, and \m{\omega/2\pi \sim 120\,\text{kHz}} for a single mode \cite{tex:supp}. Then, our model predicts the zonal speed \m{\sim 3}-\m{7 \,\text{km}/\text{s}}, which is indeed in the ballpark of recent experimental results \citep{ref:du25}. To calculate the effect of resonant absorption, one can obtain \m{\mc{Q}} for quasimonochromatic waves using \eq{eq:Qss} or, more generally, using \eq{eq:Qs}. For hot plasmas, our formulation also predicts a term proportional to \m{\omega_*} (like in the DW case above), which is consistent with \cite{ref:chen25}.

\msection{Example: RF waves} Since \eq{eq:zd} is not restricted to any particular frequency range, one can also use it to study poloidal-flow generation by RF waves. For example, based on \citep{ref:LeBlanc99}, consider \m{B \sim 3.5\,\text{T}}, \m{n \sim 10^{12}\,\text{cm}^{-3}}, \m{\tE \sim 0.2\,\text{kV}/\text{cm}}, \m{k_y \sim 40\,\text{cm}^{-1}} and \m{\omega \sim 74 \, \text{MHz}} for a linearly polarized RF wave, \eg an ion Bernstein wave (IBW). (The estimate for \m{\tE} is based on \citep{ref:jaeger00}, which reports \m{\tE \sim 0.1 \, \text{kV/cm}} in the same frequency range.) Assuming the cold limit and \m{\omega \sim \Omega_i \sim \omega \pm \Omega_i}, our model predicts the zonal speed \m{\sim k_y e^2 \tE^2/4m \Omega_i^3 \sim 3\,\text{km}/\text{s}} or larger at small \m{\omega/\Omega_i \pm 1}. Comparable speeds were indeed observed in experiments where IBWs were externally injected to study the nonresonant wave stress-induced bulk poloidal flow \cite{ref:LeBlanc99}.

\msection{Summary} We derived a compact and general formula \eq{eq:zd} for the zonal drive \m{\pd_t \Pi}  produced by any fluctuating electromagnetic, or electrostatic, field in magnetized plasma. This formula clarifies theories of AE--ZF--EP interactions that have been reported by other authors but remained difficult to interpret. In our case, \m{\pd_t \Pi} is expressed (up to local averaging or radial integration) simply through the material momentum of the fluctuating field and the rate at which field's kinetic momentum is resonantly absorbed by particles. Both are expressed through the plasma susceptibility \m{\vec{\chi}} and the field spectrum (Wigner matrix) \m{\vec{\mcc{W}}} in \eq{eq:QPmcc}. Our calculation is based on a framework (OC QLT) that has been extensively benchmarked and shown to reproduce many classic results as special cases \cite{my:ql}, and the above calculations are detailed in the Supplemental Material. Our theory also holds for higher-frequency waves, inviting one to reconsider pumping poloidal flows with externally launched RF waves for improved control of fusion plasmas.

\begin{acknowledgments}
This work was supported by the U.S.\ Department of Energy under Contract No.\ DE-AC02-09CH11466 during the time when one of the authors, IYD, worked at PPPL. IYD also thanks SJTU, his new employer, for supporting the final stage of this project.
\end{acknowledgments}


\begin{thebibliography}{10}

\bibitem{ref:rogers00}
B.~N. Rogers, W.~Dorland, and M.~Kotschenreuther, Phys. Rev. Lett. {\bf 85}, 5446 (2000).

\bibitem{ref:diamond05}
P.~H. Diamond, S.~I. Itoh, K.~Itoh, and T.~S. Hahm, Plasma Phys. Control. Fusion {\bf 47}, R35 (2005).

\bibitem{ref:zhu19}
H.~Zhu, Y.~Zhou, and I.Y. Dodin, New J. Phys. {\bf 21}, 063009 (2019).

\bibitem{ref:qiu16}
Z.~Qiu, L.~Chen, and F.~Zonca, Phys. Plasmas {\bf 23}, 090702 (2016).

\bibitem{ref:disiena21}
A.~Di Siena, T.~{G\"orler}, E.~Poli, A.~{Ba\~n\'on} Navarro, A.~Biancalani, R.~Bilato, N.~Bonanomi, I.~Novikau, F.~Vannini, and F.~Jenko, J. Plasma Phys. {\bf 87}, 555870201 (2021).

\bibitem{ref:choi23}
G.~J. Choi, P.~H. Diamond, and T.~S. Hahm, Nucl. Fusion {\bf 64}, 016029 (2023).

\bibitem{ref:gorelenkov18}
N.~N. Gorelenkov, V.~N. Duarte, M.~M.~Podesta, and H.~L. Berk, Nucl. Fusion {\bf 58}, 082016 (2018).

\bibitem{ref:gorelenkov19}
N.~N. Gorelenkov, V.~N. Duarte, C.~S. Collins, M.~Podesta, and R.~B. White, Phys. Plasmas {\bf 26}, 072507 (2019).

\bibitem{ref:gorelenkov24}
N.~N. Gorelenkov, V.~N. Duarte, M.~V. Gorelenkova, Zh. Lin, and S.~D. Pinches, Nucl. Fusion {\bf 64}, 076061 (2024).

\bibitem{ref:chen16}
L.~Chen and F.~Zonca, Rev. Mod. Phys. {\bf 88}, 015008 (2016).

\bibitem{ref:duarte17}
V.~N. Duarte, H.~L. Berk, N.~N. Gorelenkov, W.~W. Heidbrink, G.~J. Kramer, R.~Nazikian, D.~C. Pace, M.~Podesta, and M.~A.~Van Zeeland, Phys. Plasmas {\bf 24}, 122508 (2017).

\bibitem{ref:chen25}
L.~Chen, P.~Liu, R.~Ma, Z.~Lin, Z.~Qiu, W.~Wang, and F.~Zonca, Nucl. Fusion {\bf 65}, 016018 (2025).

\bibitem{ref:biancalani20}
A.~Biancalani, A.~Bottino, P.~Lauber, A.~Mishchenko, and F.~Vannini, J. Plasma Phys. {\bf 86}, 825860301 (2020).

\bibitem{ref:yan25}
Q.~Yan and P.~H. Diamond, Nucl. Fusion {\bf 65}, 116034 (2025).

\bibitem{ref:lin98}
Z.~Lin, T.~S. Hahm, W.~W. Lee, W.~M. Tang, and R.~B. White, Science {\bf 281}, 1835 (1998).

\bibitem{ref:sama24}
J.~N.~Sama \etal, Phys. Plasmas {\bf 31}, 112503 (2024).

\bibitem{ref:du25}
{X. D. {Du} \etal}, Phys. Rev. Lett. {\bf 135}, 265101 (2025).

\bibitem{ref:garcia24}
J.~Garcia \etal, Nature Comm. {\bf 15}, 7846 (2024).

\bibitem{ref:ruiz25}
J.~R.~Ruiz \etal, Phys. Rev. Lett. {\bf 134}, 095103 (2025).

\bibitem{ref:brochard24}
G.~Brochard \etal, Phys. Rev. Lett. {\bf 132}, 075101 (2024).

\bibitem{ref:brochard242}
G.~Brochard \etal, Nucl. Fusion {\bf 65}, 016052 (2024).

\bibitem{ref:barberis25}
T.~Barberis, V.~N. Duarte, E.~J. Hartigan-O'Connor, and N.~N. Gorelenkov, Nucl. Fusion {\bf 65}, 112005 (2025).

\bibitem{ref:chen12}
L.~Chen and F.~Zonca, Phys. Rev. Lett. {\bf 109}, 145002 (2012).

\bibitem{ref:qiu17}
Z.~Qiu, L.~Chen, and F.~Zonca, Nucl. Fusion {\bf 57}, 145002 (2017).

\bibitem{ref:chen22}
L.~Chen, Z.~Qiu, and F.~Zonca, Nucl. Fusion {\bf 62}, 094001 (2022).

\bibitem{ref:dewar73}
R.~L. Dewar, Phys. Fluids {\bf 16}, 1102 (1973).

\bibitem{ref:mcdonald85}
S.~W. McDonald, C.~Grebogi, and A.~N. Kaufman, Phys. Lett. A {\bf 111}, 19 (1985).

\bibitem{book:stix}
T.~H. Stix, {\it Waves in Plasmas\/} (AIP, New York, 1992).

\bibitem{my:ql}
I.~Y. Dodin, J. Plasma Phys. {\bf 88}, 905880407 (2022).

\bibitem{my:qlrmpp}
I.~Y. Dodin, Rev. Mod. Plasma Phys. {\bf 8}, 35 (2024).

\bibitem{ref:zhou19}
Y.~Zhou, H.~Zhu, and I.~Y. Dodin, Plasma Phys. Control. Fusion {\bf 61}, 075003 (2019).

\bibitem{ref:zhu21}
H.~Zhu and I.~Y. Dodin, Phys. Plasmas {\bf 28}, 032303 (2021).

\bibitem{ref:berry99}
L.~A. Berry, E.~F. Jaeger, and D.~B. Batchelor, Phys. Rev. Lett. {\bf 82}, 1871 (1999).

\bibitem{ref:batchelor99}
D.~B. Batchelor, L.~A. Berry, M.~D. Carter, and E.~F. Jaeger, International Conference on Electromagnetics in Advanced Applications, Torino, Italy (1999).

\bibitem{ref:LeBlanc99}
B.~P. LeBlanc, R.~E. Bell, S.~Bernabei, J.~C. Hosea, R.~Majeski, M.~Ono, C.~K. Phillips, J.~H. Rogers, G.~Schilling, C.~H. Skinner, and J.~R. Wilson, Phys. Rev. Lett. {\bf 82}, 331 (1999).

\bibitem{ref:elf00}
A.~G. Elfimov, G.~Amarante Segundo, R.~M.~O. Galvao, and I.~C. Nascimento, Phys. Rev. Lett. {\bf 84}, 1200 (2000).

\bibitem{ref:lin08}
Y.~Lin, J.~E. Rice, S.~J. Wukitch, M.~J. Greenwald, A.~E. Hubbard, A.~Ince-Cushman, L.~Lin, M.~Porkolab, M.~L. Reinke, and N.~Tsujii, Phys. Rev. Lett. {\bf 101}, 235002 (2008).

\bibitem{ref:guan}
X.~Guan, I.~Y. Dodin, H.~Qin, and N.~J. Fisch, Phys. Plasmas {\bf 20}, 102105 (2013).

\bibitem{ref:dodin22}
I.~Y. Dodin, J. Plasma Phys. {\bf 88}, 905880407 (2022).

\bibitem{ref:dodin24}
I.~Y. Dodin, Rev. Mod. Plasma Phys. {\bf 8}, 35 (2024).

\bibitem{ref:wong00}
H.~V. Wong, Phys. Plasmas {\bf 7}, 73 (2000).

\bibitem{ref:grebogi79}
C.~Grebogi, A.~N. Kaufman, and R.~G. Littlejohn, Phys. Rev. Lett. {\bf 43}, 1668 (1979).

\bibitem{tex:supp}
See the Supplemental Material.

\bibitem{phd:ruiz17}
D.~E. Ruiz, {\it Geometric theory of waves and its applications to plasma physics\/}, Ph.D. Thesis, Princeton University (2017), arXiv:1708.05423.

\bibitem{my:amc}
I.~Y. Dodin and N.~J. Fisch, Phys. Rev. A {\bf 86}, 053834 (2012).

\bibitem{my:sharm}
C.~Liu and I.~Y. Dodin, Phys. Plasmas {\bf 22}, 082117 (2015).

\bibitem{ref:gao06}
Z.~Gao, N.~J. Fisch, and H.~Qin, Phys. Plasmas {\bf 13}, 112307 (2006).

\bibitem{ref:gao07}
Z.~Gao, N.~Fisch, H.~Qin, and J.~R. Myra, Phys. Plasmas {\bf 14}, 084502 (2007).

\bibitem{ref:zhang26}
Y.~Zhang, T.~Adkins, M.~Barnes, A.~V. Dudkovskaia, M.~R. Hardman, P.~G. Ivanov, D.~Kennedy, and A.~A. Schekochihin, arXiv:2607.11789.

\bibitem{ref:chen24}
L.~Chen, Z.~Qiu, and F.~Zonca, Phys. Plasmas {\bf 31}, 040701 (2024).

\bibitem{ref:jaeger00}
E.~F. Jaeger, L.~A. Berry, and D.~B. Batchelor, Phys. Plasmas {\bf 7}, 3319 (2000).

\end{thebibliography}
\end{document}